\documentclass[twocolumn,showpacs,floatfix,superscriptaddress,amsmath,amssymb,prl]{revtex4}
\usepackage{amsfonts}
\usepackage{bbm}
\usepackage{mathrsfs}
\usepackage{txfonts}
\usepackage{amssymb}
\usepackage{graphicx}
\usepackage{hyperref}
\usepackage{ulem}
 \usepackage{overpic}
 \usepackage{psfrag}
\usepackage{tabularx}
\usepackage{array}
\usepackage{placeins}
\makeatletter
\renewcommand\subsection{\@startsection
{subsection}{2}{0mm}
 {-\baselineskip}
 {0.5\baselineskip}
{\FloatBarrier\normalfont\Large\bfseries}}
\makeatother
\newcommand{\be}{\begin{equation}}
\newcommand{\ee}{\end{equation}}

\newcommand{\PreserveBackslash}[1]{\let\temp=\\#1\let\\=\temp}

\newcommand{\ket} [1] {| #1 \rangle}

\begin{document}
\title{Graded Projected Entangled-Pair
State Representations and An Algorithm for Translationally Invariant
Strongly Correlated Electronic Systems on Infinite-Size Lattices in
Two Spatial Dimensions}

\author{Qian-Qian Shi} \affiliation{Centre for Modern Physics and Department of Physics,
Chongqing University, Chongqing 400044, The People's Republic of
China}

\author{Sheng-Hao Li} \affiliation{Centre for Modern Physics and Department of Physics,
Chongqing University, Chongqing 400044, The People's Republic of
China}

\author{Jian-Hui Zhao} \affiliation{Centre for Modern Physics and Department of Physics,
Chongqing University, Chongqing 400044, The People's Republic of
China}

\author{Huan-Qiang Zhou}
\affiliation{Centre for Modern Physics and Department of Physics,
Chongqing University, Chongqing 400044, The People's Republic of
China}

\begin{abstract}
An algorithm to find a graded Projected Entangled-Pair State
representation of the ground state wave functions is developed for
translationally invariant strongly correlated electronic systems on
infinite-size lattices in two spatial dimensions. It is tested for
the two-dimensional $t-J$ model at and away from half filling, with
truncation dimensions up to 6. We are able to locate a line of phase
separation, which qualitatively agrees with the results based on the
high-temperature expansions. We find that the model exhibits an
extended $s$-wave superconductivity for $J=0.4t$ at quarter filling.
However, we emphasize that the currently accessible truncation
dimensions are not large enough, so it is necessary to incorporate
the symmetry of the system into the algorithm, in order to achieve
results with higher precision.

\end{abstract}
 \maketitle

The investigation of models of strongly correlated electrons in two
spatial dimensions remains to be a major challenging issue in
condensed matter physics. Actually, no well-controlled analytical
techniques are available to study even the ground state properties,
which has led numerous theorists to appeal to numerical simulations.
Up to now, some powerful numerical approaches to classically
simulate quantum many-body lattice systems have been proposed, such
as Quantum Monte Carlo (QMC)~\cite{cep} and the Density Matrix
Renormalization Group (DMRG)~\cite{white}. However, the QMC suffers
from a notorious sign problem for both strongly correlated
electronic systems and frustrated spin systems, whereas the DMRG is
not so efficient for quantum lattice many-body systems in two
spatial dimensions.

Recently, significant advances have been made in the context of
classical simulations of quantum lattice many-body systems in terms
of the so-called Tensor Network (TN)
algorithms~\cite{vidal0,porras,nishino,Vers,vidal,qian,txiang,mera}.
These include the Matrix Product States (MPS)~\cite{Fannes} for
quantum lattice systems in one spatial dimension, the Projected
Entangled-Pair States (PEPS)~\cite{Vers}  for quantum lattice
systems in two and higher spatial dimensions, and the Multi-scale
Entanglement Renormalization Ansatz  (MERA)~\cite{mera} for quantum
lattice systems in any spatial dimensions. One of the advantages of
the TN algorithms is that, in contrast to the QMC, they do not
suffer from any sign problem, although a graded version of the TN
algorithms is necessary to take into account all signs arising from
the anti-commutivity of fermionic operators at different lattice
sites. Therefore, it is highly desirable to develop efficient graded
TN algorithms that enable us to classically simulate quantum
electronic lattice systems in two spatial dimensions. Remarkably,
algorithms to tackle signs arising from the anti-commutivity of
fermionic operators at different lattice sites for strongly
correlated electronic systems have recently been proposed in the
context of the MERA representations~\cite{fmera}.

In this paper, we develop a numerical algorithm to find a graded
Projected Entangled-Pair State (gPEPS) representation of the ground
state wave functions for translationally invariant strongly
correlated electronic systems on infinite-size lattices in two
spatial dimensions.  In our opinion, the gPEPS is a natural
extension of the PEPS to tackle quantum electronic lattice systems,
in which a parity is attached to each of the basis vectors of both
auxiliary and physical spaces that are super spaces in mathematics.
The algorithm is tested for the two-dimensional $t-J$ model at and
away from half filling, with truncation dimensions up to 6. We are
able to locate a line of phase separation (PS), which qualitatively
agrees with the results based on the high-temperature
expansions~\cite{rice}. We find that the model exhibits an extended
$s$-wave superconductivity for $J=0.4t$ at quarter filling. However,
we emphasize that the currently accessible truncation dimensions are
not large enough, so it is necessary to incorporate the symmetry of
the system into the algorithm, in order to achieve results with
higher precision.

{\it Graded PEPS representations.} Consider a translationally
invariant quantum electronic system on an infinite-size square
lattice in two spatial dimensions. Suppose it consists of the
nearest-neighbor interactions, characterized by a Hamiltonian
$H=\sum_{<ij>} h_{<ij>}$.  Our purpose is to find the ground state
wave function via an imaginary time evolution, with a randomly
chosen state as an initial state $\ket{\psi_{0}}$:
\begin{equation}\label{evo}
\ket{\psi_{\tau}}=\frac{\exp(-H\tau)\ket{\psi_{0}}}{\|\exp(-H\tau)\ket{\psi_{0}}\|},
\end{equation}
when $\tau \rightarrow \infty$, as long as the initial state is not
orthogonal to the genuine ground state.

\begin{figure}
\centering
\begin{overpic}
 [width=0.45\textwidth]{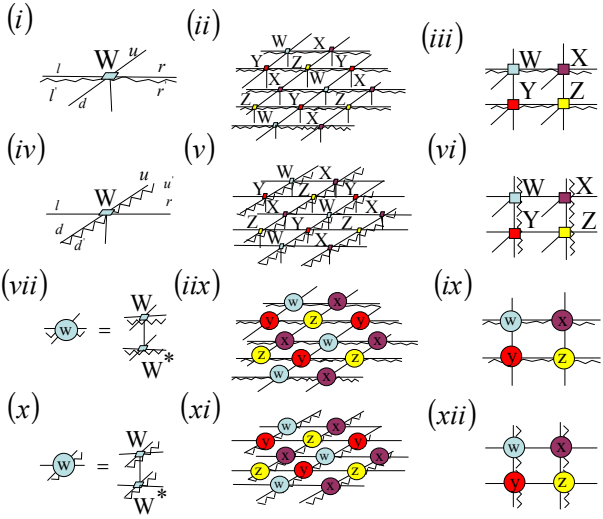}
  \end{overpic}
\caption{(color online) (i) and (iv): Seven-index tensors,
$W^{s}_{ludr;l'r'}$ and $W^{s}_{ulrd;u'd'}$ used to represent a
gPEPS representation of the system's ground state wave functions for
an infinite-size system, with $s$ being a physical index, $l$, $r$,
$u$, and $d$ denoting the inner indices. Here, $l'$, $r'$, $u'$ and
$d'$ are horizontal and vertical grading indices, respectively. (ii)
and (v): The pictorial representation of a gPEPS $\ket{\psi}$ with
horizontal and vertical grading bonds, which are used to absorb a
two-site gate acting on horizontal and vertical bonds, respectively.
(iii) and (vi): The unit cells of an infinite gPEPS with horizontal
and vertical grading bonds, respectively, made of four seven-index
tensors $W$, $X$, $Y$, and $Z$.  (vii) and (x): Double tensors
$w_{ludr;l'r'}$ and $w_{ulrd;u'd'}$ are formed from the seven-index
tensors $W$ and theirs complex conjugates $W^*$ with horizontal  and
vertical grading bonds, respectively. (viii) and (xi): The tensor
networks (TNs) for the norm of gPEPS's with horizontal and vertical
grading bonds, respectively. (ix) and (xii): The unit cells of the
TNs for the norm of gPEPS's with horizontal and vertical grading
bonds, respectively.} \label{FIG1}
\end{figure}

In order to carry out the imaginary time evolution efficiently, we
need to represent the system's ground state wave functions in terms
of graded PEPS states for translationally invariant strongly
correlated electron systems on infinite-size square lattices in two
spatial dimensions. At each site, there is a local
$\mathbbm{d}$-dimensional Hilbert super space $V$ whose basis
vectors are $|s\rangle \;(s=1,2,\cdots,\mathbbm{d})$, with the
parity $[s]$ being 0 for even vectors and 1 for odd vectors. In the
graded version of the valence bond state (gVBS)
picture~\cite{vbstate}, one may associate four
$\mathbb{D}$-dimensional auxiliary super spaces $V_l$, $V_r$, $V_u$,
and $V_d$ to the physical Hilbert super space $V$. Suppose $| l
\rangle$, $| r \rangle$, $| u \rangle$, and $| d \rangle$ are bases
of the auxiliary super spaces $V_l$, $V_r$, $V_u$, and $V_d$, with
their corresponding parities $[l], [r], [u]$ and $[d]$,
respectively. Following Ref.~\cite{Vers}, we define a gVBS state
\begin{equation}
|\Psi \rangle  =  \prod _{h,v} P_{h,v} \otimes  |\phi\rangle,
\label{vbs}
\end{equation}
where $|\phi\rangle$ is a maximally entangled state $|\phi\rangle
=\sum_{n=1}^\mathbb{D} |n,n\rangle$, the tensor product $\otimes$ is
over all possible bonds on the square lattice, and $P$ is a
projection operator $P$: $V_l \otimes V_r \otimes V_u \otimes V_d
\rightarrow V$, defined as
\begin{equation}
P = \sum _{l,y,u,d =1}^\mathbb{D} \sum _{s=1}^\mathbbm{d} W^s_{lrud}
|s\rangle \langle lrud|. \label{p}
\end{equation}
For convenience, we assume that  $W^s_{lrud} =0$ if
$[s]+[l]+[r]+[u]+[d] \neq 0 \;{\rm mod}\;2$. Substituting Eq.
(\ref{p}) into Eq.(\ref{vbs}), and taking into account the signs
arising from the grading structure, under the convention that
physical states $|s\rangle$ on a square lattice are arranged by
first ordering  from left to right along horizontal bonds and then
from up to down along vertical bonds, we may map a gVBS to a gPEPS
described by a seven-index tensor ${\tilde W}^s_{lrud;l'r'}$:
\begin{equation}
 W^s_{lrud;l'r'} = (-1)^{[r]([u]+[d])}\; (-1)^{dr'}\;
W^s_{lrud} \; \delta_{l'+r'+[u]+[d]\; {\rm mod 2},\;0}. \label{t1}
\end{equation}
Here, $l'$ and $r'$ ($l',r'=0,1$) are indices labeling two extra
horizontal grading bonds attached to each lattice site (see
Fig.~\ref{FIG1}(i)). The gPEPS for this convention is visualized in
Fig.~\ref{FIG1}(ii). However, there exists another equivalent
representation
\begin{equation}
 W^s_{lrud;l'r'} = (-1)^{[r]([u]+[d])}\;
(-1)^{ul'}\; W^s_{lrud} \; \delta_{l'+r'+[u]+[d]\; {\rm mod 2},\;0}.
\label{t2}
\end{equation}
We emphasize that, as we shall see later on,  this convention is
\textit{only} useful to absorb a two-site gate acting on a
horizontal bond during the imaginary time evolution. In order to
absorb  a two-site gate acting on a horizontal bond during the
imaginary time evolution, we need another convention that physical
states $|s\rangle$ on a square lattice are arranged by first
ordering  from up to down along horizontal bonds and then from left
to right along vertical bonds, which yields other two equivalent
representations:
\begin{equation}
W^s_{udlr;u'd'} = (-1)^{[d]([l]+[r])}\; (-1)^{rd'}\; W^s_{udlr} \;
\delta_{u'+d'+[l]+[r]\; {\rm mod 2},\;0}, \label{t3}
\end{equation}
and
\begin{equation}
 W^s_{udlr;u'd'} = (-1)^{[d]([l]+[r])}\; (-1)^{lu'}\;
W^s_{udlr} \; \delta_{u'+d'+[l]+[r]\; {\rm mod 2},\;0}. \label{t4}
\end{equation}
Here, $u'$ and $d'$ ($u',d'=0,1$) are indices labeling two extra
vertical grading bonds attached to each lattice site, as shown in
Fig.~\ref{FIG1}(iv). Note that $W^s_{udlr}$ is related to
$W^s_{lrud}$ via $W^s_{udlr} = (-1)^{([l]+[r])([u]+[d])}
W^s_{lrud}$.  The gPEPS for this convention is visualized in
Fig.~\ref{FIG1}(v).

Note that Eq. (\ref{t1}) has been introduced in Ref.~\cite{fpeps} in
the context of a fermionic PEPS (fPEPS) representation. However, an
essential difference between an fPEPS and a gPEPS lies in the fact
that the latter may be used to absorb a two-site gate during the
imaginary time evolution which acts on a horizontal bond or vertical
bond (see below). In addition, it is convenient to use super spaces
that naturally describe physical Hilbert spaces in the
two-dimensional $t-J$ model.

{\it The algorithm.} As usual, the imaginary time evolution operator
$\exp(-H\tau)$ in Eq.~({\ref{evo}}) is implemented by dividing
$\tau$ into $M$ small time slices $\delta \tau$:  $\tau = M \delta
\tau$ . For each small time slice $\delta \tau$, it is represented
by $\exp(-H\delta \tau)$. In fact, for our purpose, we shall choose
a plaquette as a unit cell, with its vertices labeled as $W,X,Y$ and
$Z$ (see Fig.~\ref{FIG1}(iii) and Fig.~\ref{FIG1}(vi)). Then, as
follows from the Suzuki-Trotter decomposition~\cite{Suzuki},
$\exp(-H\delta \tau)$ is a product of eight different kinds of
two-site gates $U_\alpha\; (\alpha
=\;WX,\;XW\;YZ,\;ZY\;WY,\;YW,\;XZ,\;ZX)$ corresponding to eight
different kinds of bonds, with the two-site gate $U_\alpha$ defined
by
\begin{equation}\label{sigate}
U_\alpha \equiv\exp(-h_{\alpha}\delta \tau), ~~\delta \tau\ll1.
\end{equation}
Thus, we have reduced the problem to implement the imaginary time
evolution to how to update the gPEPS tensors $W$, $X$, $Y$, and $Z$
under the action of a two-site gate $U_\alpha$ acting on eight
different types of bonds. An efficient (but not optimal) way to do
this is to adapt the strategy used in the iMPS
algorithm~\cite{vidal}. Therefore, we attach a diagonal singular
value matrix $\lambda_\alpha$ to each type of bonds, with tensors
$\Gamma_W$, $\Gamma_X$, $\Gamma_Y$, and $\Gamma_Z$ defined via
removing a square root of the singular value matrix from each of all
four bonds surrounding $W$, $X$, $Y$, and $Z$, respectively. As
such, the algorithm consists of two parts: first, absorb the action
of a two-site gate $U_\alpha$ on a gPEPS to update the gPEPS
tensors; second, read out the expectation value of a physical
observable in a given gPEPS.

\begin{figure}
\includegraphics[width=0.45\textwidth]{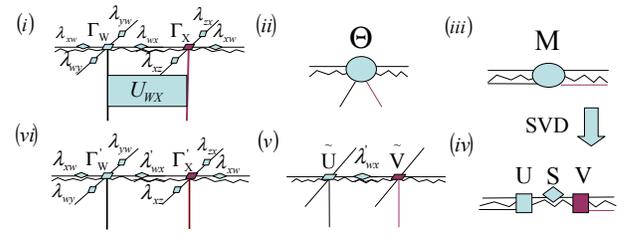}
\setlength{\abovecaptionskip}{0pt} \caption{(color online) The
procedure to update the gPEPS tensors $\Gamma_{W}$ and $\Gamma_{X}$
and the singular value matrix $\lambda_{wx}$ via absorbing the
action of a two-site gate $U_{\alpha}$. (i) the two-site gate
$U_{\alpha}$ is applied onto the gPEPS. (ii) A single tensor
$\Theta$ is formed by contracting the tensors $\Gamma_{W}$,
$\Gamma_{X}$, $\lambda_{xw}$, $\lambda_{zx}$, $\lambda_{xz}$,
$\lambda_{wy}$, $\lambda_{yw}$, and the gate $U_{\alpha}$. (iii)
Reshape the tensor $\Theta$ into a matrix $M$. (iv) A singular value
decomposition (SVD) is performed for the matrix $M$, followed by a
truncation, with only the $\mathbb{D}$ largest singular values
retained in the updated singular matrix $\lambda_{wx}^{'}$. (v)
Reshape the matrices $U$ and $V$ into the tensors {\it \~{U}} and
{\it \~{V}}. (vi) Recover the diagonal matrix $\lambda_{xw}$,
$\lambda_{zx}$, $\lambda_{xz}$, $\lambda_{wy}$, $\lambda_{yw}$, and
update the tensors $\Gamma_{W}$ and $\Gamma_{X}$ to be
$\Gamma_{W}^{'}$ and $\Gamma_{X}^{'}$.}
  \label{FIG2}
\end{figure}

(i) {\it Updating of the gPEPS tensors.} Our choice of the unit cell
in the gPEPS representation assumes that it is translationally
invariant under two-site shifts, which implies that one only needs
to address two consecutive sites linked by a certain kind of bonds;
once this is done, we simultaneously update all the tensors on the
sites linked by the same kind of bonds. The updating procedure for a
two-site gate acting on a $WX$ bond is visualized in
Fig.~\ref{FIG2}, which consists of a few steps: (i) the two-site
gate $U_{\alpha}$ is applied onto the gPEPS. (ii) A single tensor
$\Theta$ is formed by contracting the tensors $\Gamma_{W}$,
$\Gamma_{X}$, $\lambda_{xw}$, $\lambda_{zx}$, $\lambda_{xz}$,
$\lambda_{wy}$, $\lambda_{yw}$, and the gate $U_{\alpha}$. (iii)
Reshape the tensor $\Theta$ into a matrix $M$. (iv) A singular value
decomposition (SVD) is performed for the matrix $M$, followed by a
truncation, with only the $\mathbb{D}$ largest singular values
retained in the updated singular matrix $\lambda_{wx}^{'}$. (v)
Reshape the matrices $U$ and $V$ into the tensors {\it \~{U}} and
{\it \~{V}}. (vi) Recover the diagonal matrix $\lambda_{xw}$,
$\lambda_{zx}$, $\lambda_{xz}$, $\lambda_{wy}$, $\lambda_{yw}$, and
update the tensors $\Gamma_{W}$ and $\Gamma_{X}$ to be
$\Gamma_{W}^{'}$ and $\Gamma_{X}^{'}$.

\begin{figure}
\centering
 \begin{overpic}
 [width=0.45\textwidth]{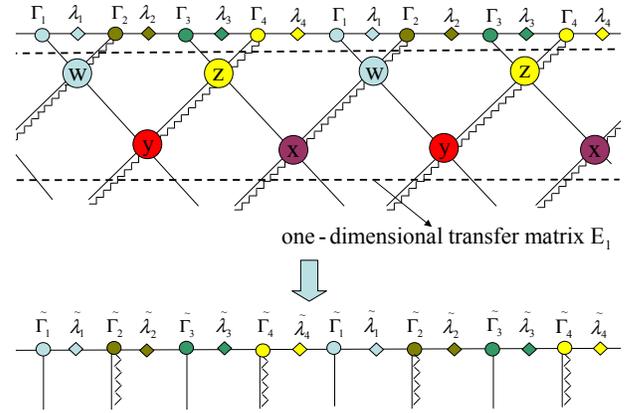}
  \end{overpic}
\caption{(color online) The iMPS used to approximate the largest
eigenvector of the one-dimensional transfer matrix $E_1$, shown here
as an Matrix Product Operator on an infinite strip. Here, we need to
absorb two-site nonunitary gate acting on an iMPS. The iMPS turns
out to be the largest eigenvector of the transfer matrix $E_1$ if
$\lambda_{1}$, $\lambda_{2}$, $\lambda_{3}$, and $\lambda_{4}$
converge after the transfer matrix $E_1$ is acted on the iMPS enough
times.}\label{FIG3}
\end{figure}

\begin{figure}
\centering
 \begin{overpic}
 [width=0.45\textwidth]{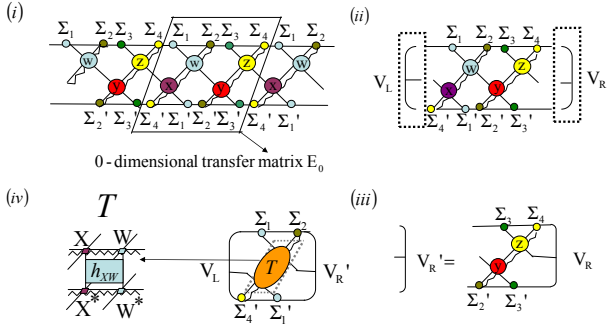}
  \end{overpic}
\caption{(color online) The ground state energy per bond is computed
by contracting the environment tensors, with the $XW$ bond as an
example. (i) The right and left largest eigenvectors of the transfer
matrix $E_1$ are denoted by tensors $\Sigma_1$, $\Sigma_2$,
$\Sigma_3$, $\Sigma_4$, and $\Sigma_1^{'}$, $\Sigma_2^{'}$,
$\Sigma_3^{'}$, $\Sigma_4^{'}$, respectively. Here, the
0-dimensional transfer matrix $E_0$ is visualized. (ii) The largest
left and right eigenvectors $V_{L}$ and $V_{R}$ of the
one-dimensional transfer matrix $E_0$. (iii)  An auxiliary vector
$V_{R}^{'}$ is defined by absorbing  the tensors $\Sigma_3$,
$\Sigma_4$, $\Sigma_2'$, $\Sigma_3'$, $y$, and $z$. (iv) The ground
state energy for the $XW$ bond is computed by contracting a tensor
$T$ with the tensors $V_L$, $V_R'$, $\Sigma_1$, $\Sigma_2$,
$\Sigma_1'$, and $\Sigma_4'$.  Here, $T$ is defined by the
Hamiltonian density $h_{XW}$ and the tensors $x$, $w$, $x^*$, and
$w^*$. }\label{FIG4}
\end{figure}

(ii) {\it Measuring a physical observable.} Once a gPEPS is
generated as a ground state wave function of a translationally
invariant quantum electronic system on an infinite-size square
lattice, we need to compute the expectation value of a physical
observable. For this purpose, the basic building blocks are double
tensors $w$, $x$, $y$ and $z$ formed from contracting the physical
indices for the gPEPS tensors $W$, $X$, $Y$, and $Z$ and their
complex conjugates (see Fig.~\ref{FIG1}(vii) and
Fig.~\ref{FIG1}(x)), respectively. As such, one may visualize the
norm for a gPEPS as a TN, as shown in Fig.~\ref{FIG1}(viii) and
Fig.~\ref{FIG1}(xi), with their unit cells plotted in
Fig.~\ref{FIG1}(ix) and Fig.~\ref{FIG1}(xii). With the double
tensors $w$, $x$, $y$ and $z$ as the building blocks, one may form
the one-dimensional transfer matrix $E_1$, which is a Matrix Product
Operator on an infinite strip (see Fig.~\ref{FIG3}). The left and
right eigenvectors corresponding to the largest eigenvalue of $E_1$
are iMPS's, from which one may form the zero-dimensional transfer
matrix $E_0$ (see Fig.~\ref{FIG4}(i)). The largest left and right
eigenvectors of $E_0$ are defined in Fig.~\ref{FIG4}(ii), which,
together with those of $E_1$, form the environment tensors. In
addition, an auxiliary vector $V_{R}^{'}$ is defined by absorbing
the tensors $\Sigma_3$, $\Sigma_4$, $\Sigma_2'$, $\Sigma_3'$, $y$,
and $z$, as visualized in Fig.~\ref{FIG4}(iii). This enables us  to
compute the ground state energy for the $XW$ bond, as shown in
Fig.~\ref{FIG4}(iv).

\begin{figure}
\centering
 \begin{overpic}
 [width=0.4\textwidth]{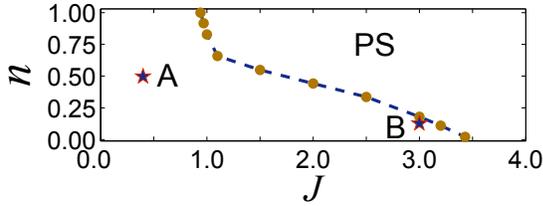}
  \end{overpic}
\caption{(color online) For $J\geq 0.95$, there is a line of phase
separation (PS). For $J\leq 0.95$, no PS occurs. Here, we have
chosen $\mathbb{D}=4$. For $J=0.4$ and $n=0.4968$, denoted as $A$,
the extended $s$-wave pairing order parameter $<\Delta>
=0.083+0.11i$, with the ground state energy per site $e=-0.9498$ for
$\mathbb{D}=6$. For $J=3.0$ and $n=0.1273$, denoted as $B$, the
extended $s$-wave pairing order parameter $<\Delta> =0.010+0.055i$,
with the ground state energy per site $e=-0.4157$ for
$\mathbb{D}=4$.}\label{FIG5}
\end{figure}
The same procedure may be used to update the gPEPS tensors and to
read out a physical observable for other bonds. However, different
conventions should be adopted for horizontal and vertical bonds.

We stress that the update procedure above is \textit{not} optimal,
in the sense that it does not produce the best approximate gPEPS
representation for each imaginary time slice during the imaginary
time evolution. As such, our update procedure can \textit{only} be
used to produce the system's ground state wave functions, but
\textit{not} for real time evolution from a prescribed initial
state. A similar situation occurs for an MPS algorithm~\cite{qian}.
This drawback may be remedied if one uses the same strategy as the
iPEPS algorithm~\cite{ipeps}, which is \textit{optimal} in the above
sense. That is, in order to absorb a two-site gate, one needs to
compute the environment tensors, i.e., the left and right largest
eigenvectors of both the one-dimensional and zero-dimensional
transfer matrices for each time slices. Therefore, the update
problem is reduced to a four-site sweep procedure that consists of
successively solving a set of linear equations~\cite{porras}.
However, this requires to update the environment tensors as we
update the gPEPS tensors $W$, $X$, $Y$, and $Z$ for each two-site
gate, so it is much less efficient.

{\it Simulation of the two-dimensional $t-J$ model.} We test the
algorithm with the two-dimensional $t-J$ model described by the
Hamiltonian~\cite{dagotto}:
\begin{equation}\label{ham}
H=-t \sum_{<ij> \sigma}[{\cal P} (c^\dagger_{i \sigma}c_{j \sigma} +
{\rm H.c.}) {\cal P}]+ J \sum_{<ij>} ({ \bf S}_i \cdot {\bf S}_j
-\frac {1}{4} n_i\;n_j),
\end{equation}
where ${ \bf S}_i$  are spin $1/2$ operators at a lattice site $i$,
${\cal P}$ is the projection operator excluding double occupancy,
and $t$ and $J$ are, respectively, the hoping constant and
anti-ferromagnetic coupling between the nearest neighbor sites
$<ij>$. Hereafter, we shall choose $t=1$ for brevity.

At half filling (i.e., $n=1$, with $n$ being the number of electrons
per site), the $t-J$ model reduces to the two-dimensional Heisenberg
model. In this case, the algorithm yields the ground state energy
per site $e=-1.1675J$,  for the truncation dimension $\mathbb{D}=4$,
quite close to the QMC simulation result
$e=-1.1680J$~\cite{mc1,mc2}. Away from half filling, the model
exhibits different behaviors for small and large anti-ferromagnetic
coupling $J$, see Fig.~\ref{FIG5}. For $J\geq 0.95$, there is a line
of PS. For $J\leq 0.95$, no PS occurs. This agrees qualitatively
with the results based on the high-temperature
expansions~\cite{rice}. Note that our result for the transition
point $J_c=3.43$ at low electron density is quite close to the exact
value $J_c= 3.4367$~\cite{hellberg}. Here, we have chosen
$\mathbb{D}=4$.

In the homogeneous regime, it turns out that the algorithm does not
yield much conclusive results, due to the fact that the truncation
dimension $\mathbb{D}$ currently accessible is quite small (up to
$\mathbb{D}=6$). However, signals of extended $s$-wave
superconductivity are observed in two regimes: the first is the
regime for $2<J<3.43$ at low electron density, and the second is the
regime which starts at least from $J=0.4$ at (almost) quarter
filling. For $J=0.4$ and $n=0.4968$, denoted as $A$ in Fig.
~\ref{FIG5}, the extended $s$-wave pairing order parameter $<\Delta>
=0.083+0.11i$, with the ground state energy per site $e=-0.9498$ for
$\mathbb{D}=6$. For $J=3.0$ and $n=0.1273$, denoted as $B$ in Fig.
~\ref{FIG5}, the extended $s$-wave pairing order parameter $<\Delta>
=0.010+0.055i$, with the ground state energy per site $e=-0.4157$
for $\mathbb{D}=4$. It remains unclear whether or not these two
points are continuously connected.

Given that the bottleneck of the algorithm to achieve higher
precision is the smallness of the truncation dimension $\mathbb{D}$,
we expect that our data may be significantly improved for a larger
truncation dimension $\mathbb{D}$ by incorporating the symmetry into
the algorithm~\cite{singh}. Indeed, even for the anti-ferromagnetic
order parameter at half filling, the currently accessible truncation
dimensions are still too small.

{\it Summary and outlook.}  We have developed a numerical algorithm
to find a gPEPS representation of the ground state wave functions
for translationally invariant strongly correlated electronic systems
on infinite-size lattices in two spatial dimensions. It is tested
for the two-dimensional $t-J$ model at and away from half filling,
with truncation dimensions up to 6. We are able to locate a line of
PS, which qualitatively agrees with the results based on the
high-temperature expansions~\cite{rice}. It is proper to stress that
the location of the line may vary if the truncation dimension is
increased, although the variation might be small (especially at low
electron density), due to the fact that PS can be seen from a
consideration based on energetics~\cite{lin}, whereas the ground
state energy per site we computed is reasonable, compared to the
exact values for a small ($4\times 4$) cluster.  In addition, the
model exhibits an extended $s$-wave superconductivity for $J=0.4t$
at quarter filling. However, we emphasize that the currently
accessible truncation dimensions are not large enough, so it is
necessary to incorporate the symmetry of the system into the
algorithm, in order to achieve results with higher precision. This
is currently under investigation.

After this work was completed, we have become aware of a preprint by
T. Barthel, C. Pineda, and J. Eisert, arXiv:0907.3689, in which an
alternative contraction scheme for the fermionic PEPS is discussed
in the context of fermionic operator circuits. This work is
supported in part by the National Natural Science Foundation of
China (Grant Nos: 10774197 and 10874252) and the Natural Science
Foundation of Chongqing (Grant No: CSTC, 2008BC2023).

\end{document}